\documentclass[aps,prl,twocolumn,amsmath,amsart,amssymb,superscriptaddress]{revtex4-2}
\usepackage[colorlinks=true, linkcolor=blue, citecolor=blue, urlcolor=blue]{hyperref}

\usepackage{graphicx}
\usepackage[T1]{fontenc}
\usepackage{amsmath}
\usepackage{color}
\usepackage{times}
\usepackage{txfonts}
\usepackage[normalem]{ulem}
\usepackage{soul}
\usepackage{xcolor}

\begin{document}
	
	
	\title{Doping evolution of spin excitations in La$_{3-x}$Sr$_{x}$Ni$_2$O$_7$/SrLaAlO$_4$ superconducting thin films}
	
	\author{Hengyang Zhong}
	\thanks{These two authors contributed equally to this work.}
	\affiliation{School of Physics and Astronomy, Beijing Normal University, and Key Laboratory of Multiscale Spin Physics (Beijing Normal University), Ministry of Education, Beijing 100875, China}

    \author{Bo Hao}
    \thanks{These two authors contributed equally to this work.}
	\affiliation{National Laboratory of Solid State Microstructures, Jiangsu Key Laboratory of Artificial Functional Materials, College of Engineering and Applied Sciences, Nanjing University, Nanjing 210093, China}
	\affiliation{Collaborative Innovation Center of Advanced Microstructures, Nanjing University, Nanjing 210093, China}
	
    \author{Anni Chen}
	\affiliation{Photon Science Division, Swiss Light Source, Paul Scherrer Institut, CH-5232 Villigen PSI, Switzerland}
	
    \author{Xinru Huang}
    \author{Chunyi Li}
    \author{Wenting Zhang}
    \author{Chang Liu}
	\affiliation{School of Physics and Astronomy, Beijing Normal University, and Key Laboratory of Multiscale Spin Physics (Beijing Normal University), Ministry of Education, Beijing 100875, China}

    \author{Yuxun Zhong}
    \affiliation{Guangdong Provincial Key Laboratory of Magnetoelectric Physics and Devices, School of Physics, Sun Yat-Sen University, Guangzhou 510275, China}

	\author{Kurt Kummer}
	\author{Nicholas Brookes}
	\affiliation{European Synchrotron Radiation Facility, BP 220, F-38043 Grenoble Cedex, France}

	\author{Dao-Xin Yao}
    \affiliation{Guangdong Provincial Key Laboratory of Magnetoelectric Physics and Devices, School of Physics, Sun Yat-Sen University, Guangzhou 510275, China}

	\author{Yuefeng Nie}
	\email{ynie@nju.edu.cn}
	\affiliation{National Laboratory of Solid State Microstructures, Jiangsu Key Laboratory of Artificial Functional Materials, College of Engineering and Applied Sciences, Nanjing University, Nanjing 210093, China}
	\affiliation{Collaborative Innovation Center of Advanced Microstructures, Nanjing University, Nanjing 210093, China}
	\affiliation{Jiangsu Physical Science Research Center, Nanjing 210093, China}

	\author{Thorsten Schmitt}
    \email{thorsten.schmitt@psi.ch}
    \affiliation{Photon Science Division, Swiss Light Source, Paul Scherrer Institut, CH-5232 Villigen PSI, Switzerland}

    \author{Xingye Lu}
    \email{luxy@bnu.edu.cn}
    \affiliation{School of Physics and Astronomy, Beijing Normal University, and Key Laboratory of Multiscale Spin Physics (Beijing Normal University), Ministry of Education, Beijing 100875, China}
	
	\date{\today}

\begin{abstract}
Ambient-pressure superconductivity in compressively strained bilayer nickelate films provides a unique platform to test pairing scenarios, yet the evolution of magnetism with carrier doping remains largely unexplored. Here, we utilize Ni $L_3$-edge resonant inelastic x-ray scattering to systematically track the evolution of spin and electronic excitations in coherently strained La$_{3-x}$Sr$_x$Ni$_2$O$_7$/SrLaAlO$_4$ thin films, spanning the superconducting ($x \le 0.21$) and overdoped non-superconducting ($x = 0.38$) regimes. We reveal that dispersive spin excitations, characterized by double-stripe correlations and nearly doping-independent exchange scales, persist robustly throughout the entire superconducting dome. In stark contrast, upon entering the overdoped non-superconducting state, this coherent magnetic framework undergoes an abrupt collapse, melting into a heavily damped, low-spectral-weight continuum. We show that this magnetic breakdown is fundamentally driven by a selective doping-induced orbital reconstruction. While the invariant $\sim\!1.0$~eV intra-atomic $dd$ peak confirms an intact local octahedral crystal field, the concurrent quenching of the $\sim\!0.4$~eV and $\sim\!1.6$~eV features signifies a severe degradation of the apical-oxygen-mediated $d_{z^2}$--$p_z$--$d_{z^2}$ singlet sector and bilayer charge-transfer coherence. The synchronized demise of coherent spin excitations and macroscopic pairing establishes a direct, doping-controlled link, underscoring that maintaining the localized $d_{z^2}$ magnetic framework and robust apical-oxygen coupling is the fundamental prerequisite for high-$T_c$ superconductivity in bilayer nickelates.

\end{abstract}

\maketitle
The discovery of hydrostatic pressure-induced superconductivity in the bilayer Ruddlesden–Popper nickelate La$_3$Ni$_2$O$_7$ ($T_{c,\rm onset}\sim80$ K) has opened a new avenue for exploring high-$T_c$ superconductivity beyond cuprates and iron-based superconductors \cite{sun2023signatures,yuan2024np,wang2024pressure,shi2025spin,Li2025}. Through isovalent rare-earth (such as Pr, Sm, Nd) substitution of La in La$_3$Ni$_2$O$_7$ \cite{wang2024bulk,li2025bulk,zhong2025evolution,qiu2025interlayer,pan2024}, the bulk superconductivity can be stabilized and further enhanced to $T_{c,\rm onset}\sim96$~K~\cite{li2025bulk}. La$_3$Ni$_2$O$_7$ consists of NiO$_2$ bilayers coupled via inner apical oxygens (O$_{\rm AP}$), hosting quasi‑degenerate Ni $3d_{x^2-y^2}$ and $3d_{z^2}$ orbitals near the Fermi level~\cite{sun2023signatures,yaodx2023,Li2025ARPES,wang2025ARPES,wang2025review,Liu2024NatCommun,yang2024orbital,Zhang2023,Christiansson2023,lechermann2023electronic,Shilenko2023,cao2024flat,Qin2023} and sizable interlayer superexchange interactions~\cite{chen2024electronic,cao2024flat,lu2024interlayer,Yang2023,shen2023effective} in a double-stripe ground state \cite{Plokhikh2025,Gupta2025,ren2025resolving,ni2025spin}. Theoretical and experimental works have emphasized the importance of interlayer antiferromagnetic (AFM) superexchange $J_z$ and spin fluctuations in driving superconductivity in this multi-orbital system~\cite{cao2024flat,lu2024interlayer,zhong2025spin,qu2024bilayer,oh2023type,Tian2024,Chen2024,Ouyang2024,Lu2025,Qu2025,sakakibara2024possible,Yang2023,shen2023effective,Qin2023,Luo2024NPJQM,Kaneko2024,Zheng2025,Kakoi2024,Wang2025}. However, the $P\gtrsim10$ GPa hydrostatic pressure severely limits systematic spectroscopic probes of the pairing interaction and complicates application-oriented characterization.

\begin{figure*}[htbp!]
	\centering
	\includegraphics[width=16cm]{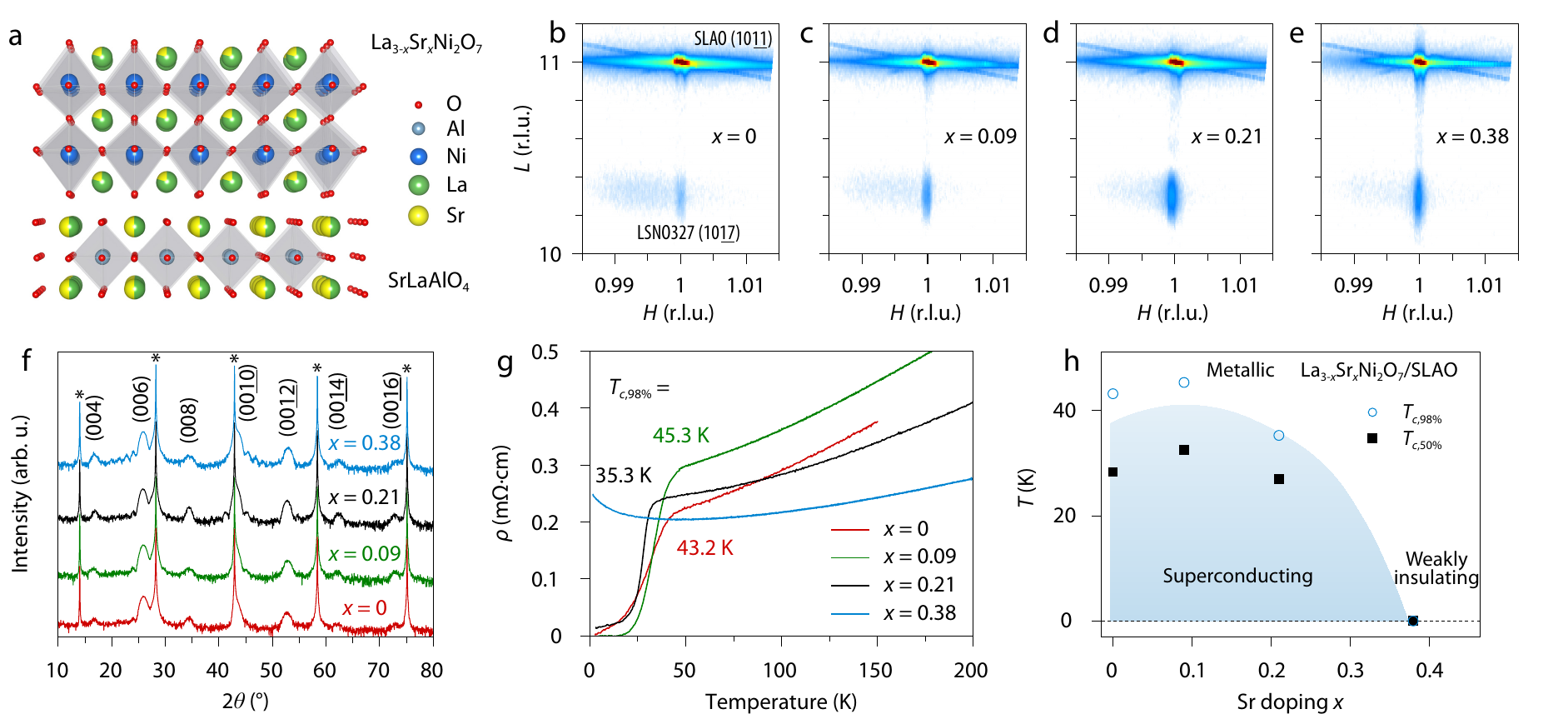}
	\caption{{\bf Epitaxial growth and superconducting phase diagram of La$_{3-x}$Sr$_x$Ni$_2$O$_7$ thin films on SrLaAlO$_4$.} \textbf{a}, Structural schematic of the La$_{3-x}$Sr$_x$Ni$_2$O$_7$ (LSNO327) film coherently grown on a (001)-oriented SrLaAlO$_4$ (SLAO) substrate. {\bf b}–{\bf e}, Reciprocal space maps in $[H,0,L]$ plane (in reciprocal lattice units) collected around the SLAO $(10 \underline{11})$ reflection showing the LSNO327 $(10 \underline{17})$ film peak for $x = 0$ (b), $0.09$ (c), $0.21$ (d) and $0.38$ (e); the vertical alignment of the film and substrate peaks indicates coherent in-plane strain. {\bf f}, X-ray diffraction $\theta-2\theta$ scans showing $(00L)$ reflections of the LSNO327 phase (indexed) for films with $x = 0, 0.09, 0.21$ and $0.38$; asterisks denote substrate-related peaks. {\bf g}, Temperature dependence of the in-plane resistivity $\rho$($T$) for representative films, highlighting superconducting transitions with $T_{c,98\%} = 43.2$ K ($x = 0$), $45.3$ K ($x = 0.09$) and $35.3$ K ($x = 0.21$), while the $x = 0.38$ film shows weakly insulating behaviour; $T_{c,98\%}$ is defined at $98\%$ of the normal-state resistivity. {\bf h}, Sr-doping evolution of the phase diagram summarizing $T_{c,98\%}$ (open circles) and $T_{c,50\%}$ (solid squares; midpoint criterion). The shaded region marks the superconducting regime.}
	\label{fig1}
\end{figure*}

Recently, compressive epitaxy on SrLaAlO$_4$(001) (SLAO), imposing an in-plane strain of $\varepsilon\approx-2\%$, has been shown to stabilize ambient-pressure superconductivity in La$_3$Ni$_2$O$_7$-based thin films with $T_{c,\rm onset}$ exceeding 40 K \cite{ko2025signatures,zhou2025ambient,liu2025nmat,hao2025nmat} and reaching $\sim$60 K in the most recent reports \cite{zhou2025superconductivity}. These epitaxial films provide experimental access to momentum- and energy-resolved spectroscopies that are central to unveiling the pairing mechanism \cite{fan2025superconducting,shen2025nodeless,sun2025observation,wang2025ARPES,zhong2025spin}. Scanning tunneling microscopy/spectroscopy on La$_2$PrNi$_2$O$_7$/SLAO reveals a fully opened superconducting gap with two characteristic energy scales and a bosonic-mode feature at $\sim$30 meV \cite{fan2025superconducting}. Complementary ARPES measurements on superconducting (La,Pr,Sm)$_3$Ni$_2$O$_7$/SLAO resolve a nodeless superconducting gap ($\Delta \approx 18$ meV) and an electron–boson coupling scale, consistent with an $s$-wave–type order parameter \cite{shen2025nodeless}. In parallel, our prior Ni $L_3$-edge RIXS study on La$_3$Ni$_2$O$_7$/SLAO (LNO/SLAO) thin films uncovered robust, dispersive magnetic excitations with collinear double-stripe correlations and enhanced interlayer exchange $J_z$ under compressive strain \cite{zhong2025spin}. Taken together, the spectroscopic evidence for a fully gapped superconducting state and strong antiferromagnetic spin fluctuations is compatible with a spin-fluctuation-mediated $s^{\pm}$ pairing scenario in bilayer nickelate films~\cite{yang2023possible,sakakibara2024possible,liu2023s,lu2024interlayer,qu2024bilayer,Tian2024,Zheng2025,gu2025effective,zhang2024structural,huang2023impurity,zhang2023trends,wang2024electronic,jiang2025}.

However, although dispersive spin excitations were observed in pure LNO/SLAO thin films, an unambiguous link between superconductivity and spin correlations remains to be established experimentally \cite{zhong2025spin}. In cuprates and iron-based superconductors, a decisive strategy to unveil this coupling is to track the evolution of spin excitations with carrier doping across the superconducting phase diagram and into the overdoped, non-superconducting regime \cite{wakimoto2007disappearance,tacon2011intense,wang2013doping,dai2015rmp}. The coherently strained La$_{3-x}$Sr$_x$Ni$_2$O$_7$/SLAO (LSNO/SLAO) thin films (Fig.~1a) provide precisely this opportunity: Sr substitution systematically tunes the carrier density and drives superconductivity from a robust regime ($x\lesssim0.21$) to an overdoped, non-superconducting state near $x\approx0.38$ under essentially fixed epitaxial strain \cite{hao2025nmat,wang2026superconducting}.

In this work, we use high-resolution Ni $L_3$-edge RIXS to track the evolution of magnetic and electronic excitations in coherently compressively strained La$_{3-x}$Sr$_x$Ni$_2$O$_7$/SLAO thin films ($x=0$, 0.09, 0.21, and 0.38), spanning the robust superconducting regime ($x\le0.21$) to an overdoped, non-superconducting state ($x=0.38$) (Fig.~1). 
With increasing $x$, the $\sim\!0.4$~eV and $\sim\!1.6$~eV electronic features, which carry strong Ni-$e_g$ character but are controlled by interlayer Ni-O${\rm AP}$-Ni charge-transfer hybridization, are progressively weakened and become largely featureless at $x=0.38$, whereas the dominant $\sim\!1.0$~eV intra-atomic $dd$ peak remains nearly unchanged. This selective evolution indicates that Sr doping preserves the fundamental local $dd$ manifold while specifically disrupting the Ni--O hybridized excitations associated with charge transfer. Fundamentally, this targeted spectral collapse could be driven by a profound doping-induced electronic reconstruction---a Lifshitz transition that severely metallizes the $d_{z^2}$ orbital and poisons the O$_{\rm AP}$-mediated interlayer hopping \cite{chen2024electronic,zhong2025spin,ryee2025optimal,wang2026superconducting}. Concomitantly, while the low-energy magnetic response remains robust and dispersive for $x\le0.21$ with undamped dispersions and a modest reduction of spectral weight, the double-stripe spin excitations abruptly soften and merge into an overdamped paramagnon continuum at $x=0.38$. This magnetic degradation reveals that the overdoped, highly itinerant $d_{z^2}$ electrons are fundamentally incompatible with the localized moments required to sustain short-range double-stripe correlations, thereby subjecting the spin dynamics to severe Landau damping. The synchronized breakdown of the coherent interlayer magnetic framework and macroscopic superconductivity at $x=0.38$ establishes a definitive, doping-controlled link: preserving the localized nature of $d_{z^2}$ electrons and their robust interlayer magnetic coupling is the fundamental prerequisite for high-$T_c$ pairing in bilayer nickelates.

\hspace*{\fill}

\noindent
\textbf{Results}

\noindent
\textbf{Sample and experimental setup}

\noindent
The La$_{3-x}$Sr$_x$Ni$_2$O$_7$ thin films investigated in this work were epitaxially grown on (001)-oriented SrLaAlO$_4$ (SLAO) substrates via reactive molecular-beam epitaxy (MBE) in a DCA R450 system~\cite{hao2025nmat}. During deposition, the substrate temperature was maintained at $\sim$720~°C under a distilled-ozone background pressure of $\sim$1$\times$10$^{-5}$~Torr. The film thickness was precisely controlled to 3 unit cells via shuttered layer-by-layer growth---monitored in situ by reflection high-energy electron diffraction (RHEED) oscillations---to prevent strain relaxation. Following growth, the samples were cooled to room temperature under the same oxidizing background to minimize oxygen vacancy formation. An \textit{ex situ} ozone-assisted annealing process (typically at $\sim$380~°C for $\sim$1~h) was subsequently performed to optimize the superconducting properties. The high crystalline quality and coherent epitaxial strain of the films were confirmed via reciprocal-space mapping (Figs.~1b--1e) and x-ray diffraction $\theta$--$2\theta$ scans (Fig.~1f) utilizing monochromated Cu $K_{\alpha1}$ radiation.

X-ray absorption spectroscopy (XAS) and RIXS measurements were conducted at the soft x-ray beamline ID32 of the European Synchrotron Radiation Facility (ESRF). All presented data were collected at a base temperature of $T \approx 20$~K. The XAS spectra were acquired in the total fluorescence yield (TFY) mode. Incident-energy-dependent RIXS spectra were recorded using $\pi$-polarized photons. Momentum-dependent RIXS maps were collected at the resonant Ni $L_3$-edge (856.4~eV) with $\pi$-polarized incident photons under a grazing-incidence geometry. These measurements were performed along two high-symmetry momentum directions, $[H, 0]$ and $[H, H]$, achieving a combined energy resolution of $\Delta E \approx 32$~meV.

Figure~1g displays the temperature-dependent in-plane resistivity $\rho(T)$ of the coherently strained LSNO/SLAO films. Evaluated using the 98\% normal-state resistivity criterion, the superconducting onset temperatures are $T_{c,98\%}=43.2$~K for $x=0$, $45.3$~K for $x=0.09$, and $35.3$~K for $x=0.21$. In striking contrast, the heavily doped $x=0.38$ film remains metallic down to $T\gtrsim 50$~K but subsequently develops a low-temperature resistivity upturn, consistent with a weakly insulating response. The doping evolution of $T_{c,98\%}$ and the midpoint transition temperature $T_{c,50\%}$ is summarized in Fig.~1h. These macroscopic transport properties map out an incomplete superconducting dome consistent with recent reports \cite{hao2025nmat,wang2026superconducting}, demonstrating that robust superconductivity persists up to $x\approx 0.21$ but is entirely extinguished down to 2~K at the critical overdoped composition of $x=0.38$.

\begin{figure}
	\centering
	\includegraphics[width=7cm]{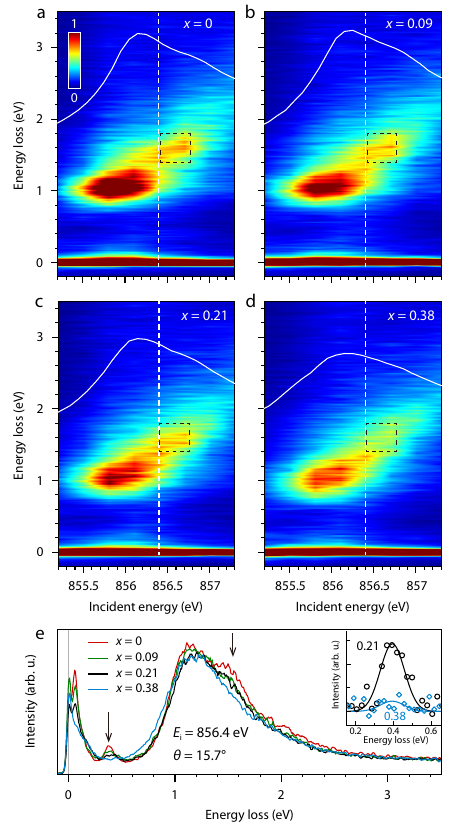}
	\caption{ {\bf Sr-doping evolution of \textit{dd} excitations in LSNO/SLAO thin films.} {\bf a}–{\bf d}, Incident-energy-dependent Ni-$L_3$ RIXS intensity maps measured at a grazing-incidence angle $\theta$ = 20$^\circ$ for films with $x=0$ ({\bf a}), 0.09 ({\bf b}), 0.21 ({\bf c}) and 0.38 ({\bf d}), plotted as a function of incident photon energy ($E_i$) and energy loss. The solid white curves show the corresponding Ni-$L_3$ X-ray absorption spectra (scaled for clarity). The black dashed squares mark the $\sim1.6$ eV excitations. The vertical dashed line marks $E_i=856.4$ eV, the incident energy used to extract the spectra in {\bf e}. {\bf e}, Comparison of Ni-$L_3$ RIXS spectra measured at $E_i=856.4$ eV and a grazing-incidence angle $\theta=15.7^\circ$ for the four dopings, highlighting the doping-dependent evolution of the excitations; vertical arrows indicate the changes of the $\sim0.4$ eV and $\sim1.6$ eV features. Data for $x=0$ are from Ref.~\cite{zhong2025spin}. The inset in {\bf e} shows a comparison of the $0.4$ eV excitation between $x=0.21$ and $0.38$.}
	\label{fig2}
\end{figure}

\hspace*{\fill}

\noindent
\textbf{Doping evolution of electronic and spin excitations}

\noindent
Figure~2 summarizes the incident-energy-dependent Ni $L_3$-edge RIXS spectra, capturing the evolution of local electronic excitations across the Sr-doping series. To understand this spectral evolution, an oxygen-inclusive description of the electronic structure is essential. Recent 11-band $d$-$p$ Hubbard model calculations show that La$_3$Ni$_2$O$_7$ films remain charge-transfer systems, with a reduced charge-transfer gap and enhanced charge-transfer capability compared with the high-pressure bulk material \cite{wu2024superexchange,zhong2026}. In this framework, the relevant low-energy electronic structure is not composed of purely atomic Ni $d$ levels, but of Ni-O hybridized sectors. Specifically, while the $\sim\!0.4$~eV mode has often been discussed in a bilayer two-orbital description as an excitation within the $e_g$ sector involving $d{x^2-y^2}$ and $d{z^2}$ states, it is intimately tied to the out-of-plane $d{z^2}$--$p_z$--$d_{z^2}$ channel mediated by the inner O$_{\rm AP}$---an important but fragile low-energy degree of freedom in the film. The $\sim\!1.0$~eV peak originates from the dominant intra-atomic $t_{2g} \rightarrow e_g$ transitions, reflecting the fundamental local octahedral crystal field. Finally, the $\sim\!1.6$~eV feature resides in a mixed $dd$/charge-transfer manifold with appreciable ligand-hole character \cite{zhong2025spin}, a transition specifically identified as charge-transfer in nature by cellular dynamical mean-field theory \cite{wu2024superexchange}.  

With these specific orbital assignments established, the distinct doping evolution of the electronic excitations naturally emerges. Within the superconducting regime ($x\le0.21$), the overall $dd$/charge-transfer manifold remains robust (Figs.~2a--2c, 2e). In this doping range, the $\sim\!0.4$~eV excitation remains resolvable as a distinct peak (albeit becoming broader with reduced contrast at $x=0.21$), and the $\sim\!1.6$~eV feature, highlighted by black dashed squares in the incident-energy-dependent maps (Figs.~2a--2d), exhibits a discernible reduction in intensity with increasing Sr content. To more clearly resolve this doping-dependent evolution, Figure~2e presents high-quality RIXS line spectra extracted at $E_i = 856.4$~eV. In these spectra, the $\sim\!1.6$~eV feature appears as a broad shoulder on the high-energy slope, as its primary resonance occurs at a slightly higher incident energy. Upon entering the overdoped $x=0.38$ state, both the $\sim\!0.4$~eV response and the $\sim\!1.6$~eV shoulder are substantially suppressed and become largely featureless.

Notably, amidst this drastic spectral suppression, the dominant intra-atomic $dd$ peak ($t_{2g} \rightarrow e_g$) at $\sim\!1.0$~eV remains virtually unchanged. This contrast indicates that the basic Ni-O octahedral crystal-field environment is largely preserved, and that the spectral collapse at $x=0.38$ reflects a selective reconstruction of specific low-energy Ni-O$_{\rm AP}$ hybridized excitations rather than a global modification of the local crystal field. Ultimately, this selective quenching of the $\sim\!0.4$~eV and $\sim\!1.6$~eV features provides a direct spectroscopic fingerprint of a profound doping-induced electronic reconstruction \cite{wu2024superexchange,zhong2026,ryee2025optimal}.

This picture naturally accounts for the distinct doping evolution of the electronic excitations. As Sr-doping increases, the rapid accumulation of holes in the oxygen ligands effectively Pauli-blocks these active pathways \cite{zhong2026}, leading directly to the quenching of the 1.6~eV spectral weight. Simultaneously, although the $d_{z^2}$ orbital itself is less directly depleted of electrons, the global shift in chemical potential inevitably drives the $d_{z^2}$-dominated $\gamma$ pocket across the Fermi level---a topological Lifshitz transition \cite{ryee2025optimal}. More importantly, in La$_{3-x}$Sr$_x$Ni$_2$O$_7$ films, hole doping is expected to affect not only the in-plane $d_{x^2-y^2}$--$p_{x,y}$ sector but also the out-of-plane $d_{z^2}$-$p_z$ sector \cite{zhong2026}. Thus, the O$_{\rm AP}$-assisted bilayer channel is directly renormalized by Sr doping. At $x=0.38$, the enhanced carrier density, reduced charge-transfer gap, and direct doping of the out-of-plane Ni-O$_{\rm AP}$ sector provide efficient decay channels for this excitonic mode, causing the well-defined 0.4~eV charge-transfer excitation to melt into a broad particle-hole continuum.
In contrast, the invariance of the $\sim\!1.0$~eV peak confirms that the fundamental Ni-O octahedral crystal field is preserved. Together, the selective quenching of these two specific features provides a clear spectroscopic fingerprint of a profound, doping-induced orbital reconstruction.

\begin{figure}[htbp!]
	\centering
	\includegraphics[width=7cm]{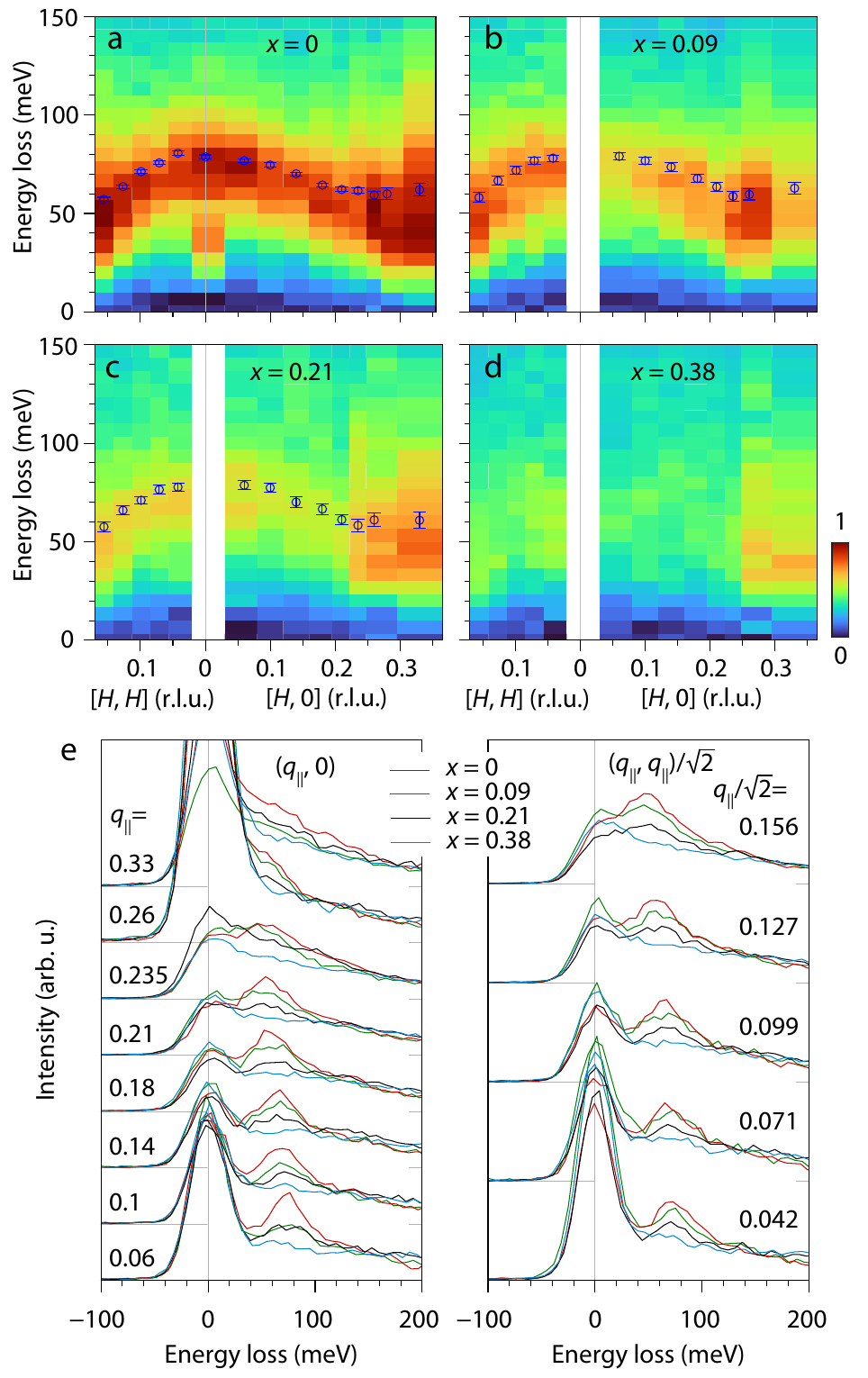}
	\caption{{\bf Sr-doping evolution of spin excitations in LSNO/SLAO thin films.} {\bf a}–{\bf d}, False-colour maps of the low-energy Ni–$L_3$ RIXS intensity plotted as a function of energy loss and in-plane momentum transfer $\mathbf{q}_\parallel$ along the $[H,H]$ direction (left part of each panel) and the $[H,0]$ direction (right part), for $x=0$ ({\bf a}), 0.09 ({\bf b}), 0.21 ({\bf c}) and 0.38 ({\bf d}). Elastic scattering has been fitted and subtracted from the color maps. The scattering angle $2\theta_s= 110^\circ$ for $|\mathbf{q}_\parallel|\ge0.26$ and $2\theta_s= 90^\circ$ otherwise.
	The symbols overlaid on the maps denote the undamped spin-excitation dispersion $E_0(\mathbf{q}_\parallel)$ extracted from fits to the RIXS spectra (error bars indicate fitting uncertainties). {\bf e}, Waterfall plots of representative low-energy Ni–$L_3$ RIXS spectra measured with $\pi$-polarized incident X-rays at $T\approx 20$ K for $x=0$, 0.09, 0.21 and 0.38, along ($\mathbf{q}_\parallel,0$) (left) and $(\mathbf{q}_\parallel,\mathbf{q}_\parallel)/\sqrt{2}$ (right); spectra are vertically offset for clarity and $\mathbf{q}_\parallel$ values are indicated. Dispersive, well-defined spin excitations persist up to $x=0.21$, whereas for $x=0.38$ the magnetic response is strongly broadened and its intensity is markedly reduced over the entire measured $\mathbf{q}_\parallel$ range. Data for $x=0$ are from Ref.~\cite{zhong2025spin}.}

	\label{fig3}
\end{figure}

Having established that Sr doping progressively weakens and broadens the $\sim\!0.4$ and $\sim\!1.6$~eV orbital-excitation features \cite{wu2024superexchange,zhong2026,ryee2025optimal}, we now turn to the low-energy magnetic response. To directly track how spin correlations evolve across a doping-tuned thin-film phase diagram at essentially fixed epitaxial strain, we performed momentum-resolved Ni $L_3$-edge RIXS measurements ($\pi$ polarization, grazing-incidence geometry; $\Delta E\approx32$~meV) on La$_{3-x}$Sr$_x$Ni$_2$O$_7$/SLAO films at the same Sr concentrations as in Fig.~2 ($x=0$, 0.09, 0.21, and 0.38). The scattering angle was set to $2\theta_s=90^\circ$ and $110^\circ$ to reduce elastic scattering. These measurements provide a direct basis for comparing the coherence, dispersion, and spectral weight of collective spin excitations between the superconducting and overdoped non-superconducting regimes (Fig.~3).

Figure~3 provides a direct visualization of the Sr-doping evolution of spin excitations in the LSNO/SLAO films. The RIXS intensity maps (Figs.~3a--3c) reveal a well-defined, dispersive magnetic mode for all superconducting compositions ($x=0$, 0.09, and 0.21) along both the $[H, H]$ and $[H, 0]$ high-symmetry directions. For $x=0$, the spectrum is dominated by a coherent collective magnetic mode that reaches its maximum energy near the zone center $\Gamma$ and disperses downward to a minimum around $\mathbf{q}_\parallel \approx (1/4, 1/4)$, consistent with robust collinear (double-stripe-like) spin correlations \cite{zhong2025spin}. Crucially, for $x=0.09$ and $0.21$ (Figs.~3b and 3c), these dispersive features remain clearly resolved. The visual dispersion boundaries are essentially unchanged compared with $x=0$, indicating that the underlying collinear double-stripe spin correlations and their characteristic exchange scale persist across the superconducting regime, suffering only a modest reduction in spectral weight.

In sharp contrast, the overdoped, non-superconducting $x=0.38$ film (Fig.~3d) exhibits a qualitatively different magnetic response. The excitation spectrum becomes strongly broadened and continuum-like, and the dispersive peak is no longer clearly resolved over the measured $\mathbf{q}_\parallel$ range, accompanied by a pronounced loss of spectral weight. This abrupt doping evolution is most transparent in the $\mathbf{q}_\parallel$-dependent waterfall spectra (Fig.~3e). For $x\le0.21$, the spectra display pronounced magnon-like peaks that coherently disperse with momentum, whereas for $x=0.38$ the response is dominated by a broad, diffuse intensity with completely quenched peak contrast. Taken together, these raw data directly show that the coherent double-stripe spin excitations are robust throughout the superconducting compositions but phenomenologically collapse into an incoherent continuum in the overdoped non-superconducting film, setting the stage for a rigorous quantitative analysis.

\begin{figure}
	\centering
	\includegraphics[width=8cm]{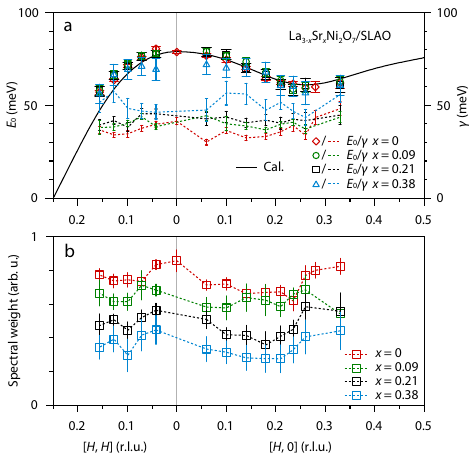}
	\caption{ {\bf Dispersions, damping and spectral weight of spin excitations across Sr doping in LSNO/SLAO thin films.} {\bf a}, Undamped spin-excitation energy $E_0(\mathbf{q}_\parallel)$ (symbols; left axis) and damping factor $\gamma(\mathbf{q}_\parallel)$ (data points connected by dashed lines; right axis) extracted from damped-harmonic-oscillator (DHO) fits to the low-energy Ni–$L_3$ RIXS spectra for $x=0$, $0.09$, $0.21$ and $0.38$, plotted along the $[H, H]$ (left) and $[H, 0]$ (right) directions. The vertical grey line marks $\Gamma$ ($\mathbf{q}_\parallel=0$), and the solid black curve shows the fitted dispersion for comparison. For $x\le 0.21$, the dispersions nearly coincide and the modes remain weakly damped, whereas for $x=0.38$ the dispersion softens and the damping increases markedly over the full $\mathbf{q}_\parallel$ range. {\bf b}, Momentum dependence of the magnon spectral weight $W(\mathbf{q}_\parallel)$, obtained by energy integration of the fitted DHO lineshapes. Error bars in {\bf a, b} represent uncertainties from the fits; dashed lines are guides to the eye.
	 }
	\label{fig4}
\end{figure}

To quantify this doping evolution of the spin excitations, we use a damped-harmonic-oscillator (DHO) response function to fit the low-energy part of each Ni $L_3$-edge RIXS spectrum and extract the momentum-dependent parameters \cite{lu2022}:
\begin{equation}
S(\mathbf{q},E)=A\,\frac{4\gamma E E_0}{(E^2-E_0^2)^2+(2\gamma E)^2},
\end{equation}
where $E_0(\mathbf{q})$ is the undamped mode energy and $\gamma(\mathbf{q})$ characterizes the damping (linewidth). The prefactor $A$ sets the overall intensity of the magnetic response; the magnetic spectral weight is obtained from the integrated area of the fitted DHO component (within the chosen energy window), while $E_0(\mathbf{q})$ directly yields the fitted dispersion.

Figure~4 quantifies the doping evolution of the magnetic mode by tracking the extracted undamped dispersion $E_0(\mathbf{q}_\parallel)$ and damping $\gamma(\mathbf{q}_\parallel)$. For the superconducting compositions measured here ($x=0$, 0.09, and 0.21), the RIXS spectra exhibit well-defined magnon-like peaks, making the extracted $E_0$ and $\gamma$ highly reliable. The dispersions for $x\le 0.21$ are essentially coincident along both $[H, H]$ and $[H, 0]$, providing a quantitative confirmation that the characteristic double-stripe spin correlations and their underlying exchange scale persist across this high-$T_c$ portion of the superconducting dome. Consistent with our previous double-stripe analysis \cite{zhong2025spin}, the dispersion can be described by linear spin-wave calculations of a classical Heisenberg model $H=\sum_{i<j}J_{ij}\mathbf{S}_i\cdot\mathbf{S}_j$ (SpinW), yielding exchange parameters $SJ_1 \approx -8.0$~meV, $SJ_2 \approx -3.7$~meV, and a dominant interlayer coupling $SJ_z \approx 44.4$~meV for $x=0$--$0.21$ (black curve in Fig.~4a). The corresponding damping remains modest in this doping range (Fig.~4b), consistent with well-defined collective spin excitations in the superconducting compositions.

For the overdoped, non-superconducting $x=0.38$ film, the magnetic response is strongly broadened and continuum-like, so extracting a unique collective-mode dispersion is intrinsically less constrained. To estimate an \textit{upper bound} on any remaining coherent component, we fit the spectra using the same DHO form (with a constrained bimagnon continuum) and test its consistency with the double-stripe Heisenberg description. Under this procedure, Fig.~4a indicates a modest softening of the band top by $\sim10-12$~meV, corresponding to reduced effective exchanges $SJ_1 \approx -5.5$~meV, $SJ_2 \approx -2.6$~meV, and $SJ_z \approx 35.2$~meV, together with a pronounced enhancement of $\gamma(\mathbf{q}_\parallel)$ relative to $x\le 0.21$ (Fig.~4b). Meanwhile, the integrated magnetic spectral weight decreases smoothly with Sr content and is $\sim$50\% lower at $x=0.38$ than at $x=0$. Given the lack of a clearly resolved dispersive peak in Fig.~3d, these $x=0.38$ parameters should be viewed as a highest estimate; the data are fully consistent with the coherent (magnon-like) component being essentially quenched, leaving a predominantly incoherent continuum.

The selective reconstruction of the Ni--O hybridized electronic excitations discussed above provides a natural microscopic context for the strong broadening of magnetic excitations at $x=0.38$. For the superconducting compositions ($x\le0.21$), the apical-oxygen-mediated $d_{z^2}$--$p_z$--$d_{z^2}$ pathway remains sufficiently coherent to support sizable interlayer magnetic coupling, while the in-plane $d_{x^2-y^2}$--$p_{x,y}$ sector remains itinerant enough to accommodate doped carriers. The modest reduction of magnetic spectral weight across this doping range indicates that carrier doping weakens but does not destroy the short-range double-stripe correlations. This stability suggests that superconductivity in the film regime is compatible with a delicate coexistence of itinerant carriers and robust local/interlayer spin correlations, rather than requiring a purely localized $d_{z^2}$ sector\cite{wu2024superexchange,zhong2026,ryee2025optimal}.

The overdoped non-superconducting film ($x=0.38$) represents a regime where this balance is lost. The suppression of the $\sim\!0.4$~eV excitation associated with the out-of-plane $d_{z^2}$--$p_z$--$d_{z^2}$ singlet sector, together with the disappearance of the $\sim\!1.6$~eV mixed $dd$/charge-transfer feature, indicates that the O$_{\rm AP}$-assisted Ni--O charge-transfer channel and bilayer electronic coherence are strongly degraded. In such a reconstructed electronic background, spin excitations can decay efficiently into low-energy particle-hole continua associated with the doped Ni--O hybridized bands. This enhanced damping, together with a reduction of the effective exchange scale, naturally accounts for the observed transformation from coherent magnon-like modes to a broad, low-spectral-weight continuum. Thus, the collapse of the magnetic response at $x=0.38$ should be understood not simply as the consequence of a rigid-band Lifshitz transition or complete delocalization of $d_{z^2}$ electrons\cite{ryee2025optimal,wang2026superconducting}, but as the combined effect of reduced charge-transfer coherence, direct doping of the out-of-plane Ni-O$_{\rm AP}$ sector, and enhanced itinerant decay channels.

\hspace*{\fill}

\noindent
\textbf{Discussion}

\noindent
In cuprates and iron-based superconductors, magnetic excitations span a broad hierarchy of energies and encode complementary aspects of the pairing problem. The high-energy spin excitations (paramagnons/spin waves) mainly reflects the persistence of spin correlations ($J$), which is suggested to serve as a pairing interaction in spin-fluctuation pairing theories \cite{wakimoto2007disappearance,tacon2011intense,wang2013doping,dai2015rmp}. The low-energy sector is strongly shaped by itinerant electrons and often shows the most direct fingerprints of superconductivity, including the appearance of a superconductivity-induced spin-resonance mode and a pronounced doping dependence of the low-energy spectral weight \cite{dai2015rmp,wakimoto2004direct,wang2013doping,Scalapino2012}. Tracking how these spin excitations evolve across the superconducting dome, especially into the overdoped regime where superconductivity vanishes, has therefore been a decisive experimental strategy for establishing a link between magnetism and pairing in both families~\cite{wakimoto2007disappearance,tacon2011intense,wang2013doping,dai2015rmp}.

Our results place Sr-doped bilayer nickelate films squarely into this comparative framework, yet reveal a distinctive outcome. In the superconducting compositions ($x \le 0.21$), the collective magnetic mode remains well-defined and strongly dispersive, and the underlying double-stripe exchange scale is largely preserved. In the overdoped non-superconducting film ($x=0.38$), however, the magnetic response becomes strongly broadened and loses substantial spectral weight. This behaviour contrasts with the persistence of robust high-energy paramagnons often observed in heavily overdoped cuprates \cite{tacon2011intense,lu2021magnetic}, and points to a stronger sensitivity of bilayer nickelate magnetism to the coherence of the apical-oxygen-mediated Ni--O network.

The accompanying evolution of the electronic excitations provides the microscopic origin for this unique magnetic sensitivity. As established earlier, the persistence of the $\sim\!1.0$~eV intra-atomic $dd$ peak confirms that the local octahedral crystal-field environment is not globally destroyed by Sr overdoping. Instead, the simultaneous collapse of the $\sim\!0.4$~eV and $\sim\!1.6$~eV features at $x=0.38$ reflects a targeted disruption of the apical-oxygen-mediated $d_{z^2}$--$p_z$--$d_{z^2}$ singlet sector and the broader bilayer charge-transfer coherence. This loss of electronic coherence is the direct spectral manifestation of the extreme metallization of the $d_{z^2}$-dominated states, often associated with a doping-induced Lifshitz transition. Ultimately, this confirms that extreme overdoping does not merely shift non-interacting energy levels, but forces these states into a highly itinerant regime, fundamentally dismantling the coherent Ni--O charge-transfer network required to sustain robust interlayer spin correlations.

The synchronized suppression of coherent spin excitations, apical-oxygen-assisted charge-transfer excitations, and superconductivity at $x=0.38$ establishes a direct doping-controlled link between Ni--O electronic coherence and pairing. Rather than implying that superconductivity requires a strictly localized $d_{z^2}$ orbital, our data suggest that the high-$T_c$ phase relies on maintaining sufficient coherence of the out-of-plane Ni--O charge-transfer channel and the associated interlayer magnetic correlations while accommodating itinerant carriers in the broader $e_g$ manifold.

Looking forward, bilayer nickelate films offer an unusually favorable geometry for momentum-resolved ``phase-diagram spectroscopy'' with soft x-ray RIXS. The characteristic double-stripe wave vector near $\mathbf{q}\!\approx\!(1/4,1/4)$ lies well within the momentum window of Ni $L_3$ RIXS, making La$_3$Ni$_2$O$_7$-based films a rare system among high-$T_c$ superconductors, in which both the ordering wave vector and its collective excitations can be mapped directly at the corresponding transition-metal $L_3$ edge. With the new generation RIXS with energy resolution approaching $\sim 10$ meV, it should become feasible to search for a superconductivity-induced spin resonance in the $\sim 15-25$ meV range (using the empirical scaling $E_r\sim4-6\,k_B T_c$, which gives $E_r\!\sim\!20$ meV for $T_c\!\sim\!50$ K)~\cite{Scalapino2012}. Detecting such a mode and tracking its intensity with Sr doping alongside the damping and spectral weight trends established here would provide an exceptionally direct test of sign-changing pairing (e.g., $s^\pm$) and offer unprecedented constraints on the pairing interaction in bilayer nickelate superconductors \cite{eremin2024}.

{\bf Author contributions}

X.L., T.S., and Y.N. conceived this project. B.H. grew the thin films and did the transport and x-ray diffraction measurements. H.Z., A.C, X.H., C.Li., W.Z, C.Liu. and X.L. performed the RIXS experiments with the help from K.K. and N.B.. H.Z., A.C., and X.L. analysed the data. Y.Z.and D.-X.Y. conducted the charge transfer analysis.
X.L. wrote the manuscript with inputs from H.Z., B.H., D.-X.Y., Y.N. and T.S.. All authors made comments.

{\bf Acknowledgement}

We would like to thank Xinman Ye for the helpful discussion. The work is supported by the Scientific Research Innovation Capability Support Project for Young Faculty (ZYGXQNJSKYCXNLZCXM-M2), the National Natural Science Foundation of China (Grants no. 12574142, 12434002, 12494591, 92565303), National Key Projects for Research and Development of China with Grant No. 2021YFA1400400, Natural Science Foundation of Jiangsu Province (No. BK20233001) and Guangdong Provincial Quantum Science Strategic Initiative (GDZX2401010). The work at PSI is supported by the Swiss National Science Foundation through project No. 207904. We acknowledge the European Synchrotron Radiation Facility (ESRF) for providing synchrotron radiation facilities under proposal numbers SC-5699 at the ID32 beamline.

\end{document}